\documentclass[sigconf]{acmart}

\AtBeginDocument{%
  \providecommand\BibTeX{{%
    Bib\TeX}}}

\copyrightyear{2024}
\acmYear{2024}
\setcopyright{rightsretained}
\acmConference[MOBIHOC '24]{The Twenty-fifth International Symposium on Theory, Algorithmic Foundations, and Protocol Design for Mobile Networks and Mobile Computing}{October 14--17, 2024}{Athens, Greece}
\acmBooktitle{The Twenty-fifth International Symposium on Theory, Algorithmic Foundations, and Protocol Design for Mobile Networks and Mobile Computing (MOBIHOC '24), October 14--17, 2024, Athens, Greece}\acmDOI{10.1145/3641512.3690165}
\acmISBN{979-8-4007-0521-2/24/10}

\acmConference[MobiHoc '24]{}{October 14--17, 2024}{Athens, Greece}

\usepackage{amsmath,amsfonts}
\usepackage{algorithmic}
\usepackage{graphicx}
\usepackage{textcomp}
\usepackage{xcolor}
\def\BibTeX{{\rm B\kern-.05em{\sc i\kern-.025em b}\kern-.08em
    T\kern-.1667em\lower.7ex\hbox{E}\kern-.125emX}}

\usepackage{mathtools,nccmath,bbold}

\DeclareMathOperator*{\argmin}{\arg\!\min}
\DeclareMathOperator{\E}{E}

\usepackage{algorithm}
\usepackage{array}
\usepackage[caption=false,font=normalsize,labelfont=sf,textfont=sf]{subfig}
\usepackage{stfloats}
\usepackage{verbatim}
\usepackage{enumitem}
\usepackage[normalem]{ulem}
\usepackage{cancel}
\usepackage{tikz}
\usetikzlibrary{matrix}
\usepackage{tikz-network}
\usetikzlibrary{chains,shapes.multipart}
\usetikzlibrary{shapes,calc,fit}
\usetikzlibrary{automata,positioning}

\theoremstyle{acmdefinition}
\newtheorem{remark}{Remark}
\newtheorem{definition}{Definition}

\begin{document}

\title{Variable-Length Stop-Feedback Coding for Minimum Age of Incorrect Information}


\author{Konstantinos Bountrogiannis}
\affiliation{%
    \institution{University of Crete}
    \city{Heraklion}
    \country{Greece}}
\email{kbountrogiannis@csd.uoc.gr}

\author{Ioannis Papoutsidakis}
\affiliation{%
    \institution{University of Bristol}
    \city{Bristol}
    \country{UK}}
\email{ioannis.papoutsidakis@bristol.ac.uk}

\author{Anthony Ephremides}
\affiliation{%
    \institution{University of Maryland}
    \city{College Park, MD}
    \country{USA}}
\email{etony@umd.edu}

\author{Panagiotis Tsakalides}
\affiliation{%
    \institution{University of Crete}
    \city{Heraklion}
    \country{Greece}}
\email{tsakalid@csd.uoc.gr}

\author{George Tzagkarakis}
\affiliation{%
    \institution{Foundation for Research and Technology -- Hellas}
    \city{Heraklion}
    \country{Greece}}
\email{gtzag@ics.forth.gr}

\settopmatter{printacmref=false} 
\renewcommand\footnotetextcopyrightpermission[1]{}

\begin{abstract}
The Age of Incorrect Information (AoII) is studied within the context of remote monitoring a Markov source using variable-length stop-feedback (VLSF) coding. Leveraging recent results on the non-asymptotic channel coding rate, we consider sources with small cardinality, where feedback is non-instantaneous as the transmitted information and feedback message have comparable lengths. We focus on the feedback sequence, i.e. the times of feedback transmissions, and derive AoII-optimal and delay-optimal feedback sequences. Our results showcase the impact of the feedback sequence on the AoII, revealing that a lower average delay does not necessarily correspond to a lower average AoII. We discuss the implications of our findings and suggest directions for coding scheme design.
\end{abstract}

\ccsdesc[500]{Networks~Network performance analysis}
\ccsdesc[500]{Networks~Network performance modeling}
\ccsdesc[500]{Mathematics of computing~Coding theory}
\ccsdesc[500]{Mathematics of computing~Information theory}
\ccsdesc[500]{Mathematics of computing~Mathematical optimization}

\keywords{Age of Incorrect Information (AoII), Finite Blocklength, Feedback Sequence}

\maketitle

\renewcommand{\thefootnote}{}
\footnotetext{This is the extended version of the paper presented at the MobiHoc '24 Workshops.\\
Published version: \url{https://doi.org/10.1145/3641512.3690165}}
\renewcommand{\thefootnote}{\arabic{footnote}}

\section{Introduction}\label{sec:introduction}

A noteworthy portion of the recent literature has been devoted to low-latency communications and their impact on modern applications. Even though the holy grail of this family of problems is the delay characterization of a network \cite{10121104}, there are special applications where the timeliness of the data is the main requirement. The increasing list of examples includes autonomous driving, critical infrastructure monitoring, emerging augmented reality network applications, and haptic communications. This aspect of communication is captured by the Age of Information (AoI) metric, which evaluates the freshness of the status updates received at a monitor from a remote source \cite{6195689}. More specifically, the instantaneous AoI at time $t$ is defined as the difference $t-u_t$, where $u_t$ is the time-stamp of the most recent update.

Since its introduction, AoI has attracted the interest of researchers and engineers from many fields~\cite{bib:aoi_survey}. Nevertheless, a shortcoming of the conventional AoI metric is that it quantifies the information freshness but omits the dynamics of the data source. For example, consider a source that changes rapidly and another that changes slowly. After some time, the samples generated simultaneously from the two sources will have the same AoI, yet the sample of the rapidly changing source is possibly less useful for decision-making.

\begin{figure}
    \centering
    \includegraphics[width=.85\linewidth]{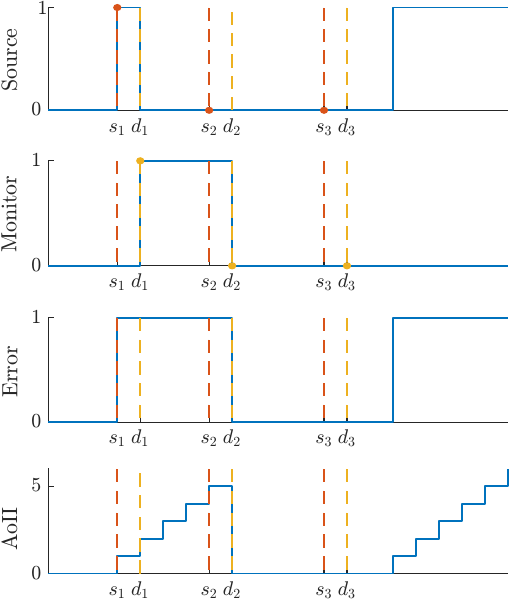}
    \vspace{-5pt}
    \caption{An example where a binary {source} is sampled and transmitted at every time slot. Successful decodings occur at time slots $d_i$, $i=1,2,3$, whereas the respective samples were generated at slots $s_i=d_i-1$. The {error} describes the mismatch between the {source} and the received values at the {monitor}. The {AoII} measures the time the {error} has been positive.}
    \label{fig:aoi_vs_aoii}
\end{figure}

This observation led to the proposal of the Age of Incorrect Information (AoII) metric, which is the main focus of this work. AoII is a content-aware metric that measures the time that information at the monitor is incorrect, weighted by the magnitude of this incorrectness as measured by an error function \cite{9137714}. More precisely, let $g(X_t,\hat{X_t})$ denote an error function at time $t$ between the source $X_t$ and its estimation $\hat{X_t}$ at the receiver. Moreover, define the age function,

\begin{equation}\label{eq:aoi}
    \Delta_t \triangleq t-h_t\,,
\end{equation}
where $h_t$ is the last time instant when the error $g(X_{h_t},\hat{X_{h_t}})$ was zero. The instantaneous AoII at time $t$ is simply the product
\begin{equation}\label{eq:aoii_def}
    \delta_t \triangleq \Delta_t \cdot g(X_t,\hat{X_t}).
\end{equation}
An example of the AoII metric is shown in Fig.~\ref{fig:aoi_vs_aoii}.

The AoII has been studied in a system with a Markov source, a constant transmission delay and a resource constraint in~\cite{9137714}, and with task-oriented age functions in~\cite{bib:aoii_semantics1}. In~\cite{bountrogiannis2023age}, the setup is extended with HARQ for error correction. In~\cite{kriouile2022minimizing}, the authors consider the case where the Markov source parameters are unknown. In~\cite{10225853}, a system with arbitrarily distributed transmission delay is considered. 

The above works assume communication over an erroneous channel, where the probability of error and transmission time are taken as fixed parameters. However, these two parameters are correlated since a lower probability of error requires a longer blocklength. Moreover, these parameters are channel- and source-dependent. Combining information-theoretic results and timeliness metrics has been a major challenge since most classical results are asymptotic, assuming infinite blocklength. Recently, non-asymptotic achievability and converse bounds were proved for fixed blocklength codes in~\cite{poly} and their variable-length counterparts in~\cite{5961844}.

Our work analyzes the minimum achievable average AoII in the non-asymptotic regime, where we investigate the impact of feedback time instances for variable-length stop-feedback (VLSF) codes. With VLSF coding, information is encoded into a theoretically infinite codeword, which is then segmented into packets. These packets are sequentially transmitted over the channel. Upon receiving a packet, the receiver attempts to decode the information, taking into consideration all previously received packets. Suppose the decoding is successful, meaning that the probability of error is less than or equal to a specified constant $\epsilon$. In that case, the receiver notifies the transmitter with an acknowledgement (ACK) feedback message. Otherwise, a negative acknowledgement (NACK) is sent to request more channel outputs. Noticeably, this communication setting is very similar to hybrid automatic repeat request with incremental redundancy (HARQ-IR).

We consider sources with small cardinality, which implies that the transmitted information and feedback message may have similar lengths. Hence, we assume that feedback is not instantaneous, thus delaying the transmission of additional coding symbols when needed by the decoder. 
We are interested in optimizing the \emph{feedback sequence}, i.e. the time slots where decoding occurs and feedback is generated. Essentially, the feedback sequence determines the length of the individual packets that a codeword is segmented into.

To derive optimal feedback sequences, it is necessary to use the probability mass function of the required blocklength for successful decoding. Since there is no closed-form result for VLSF codes, we approximate it. More precisely, we use Monte Carlo methods that simulate variable-length transmissions by calculating a bound on the probability of error and terminating the transmission when the probability of error achieves a threshold $\epsilon$. This method was introduced in~\cite{5961844} for the binary symmetric channel and later iterated for the Gaussian channel in~\cite{papouvlsf}. Our approach works with any preferred probability mass function, but we utilize the approximation for the Gaussian channel since it is a highly important model for wireless communications.

Several works have dealt with variable-length coding with finite blocklength in the context of the AoI metric. Yet, most of them utilize the fixed-blocklength results of~\cite{poly} in a simple ARQ setting assuming the decoder has perfect knowledge of decoding errors. Examples of such works include~\cite{8445909} which analyzes variable-length codes, the paper~\cite{9126228} that approximates the optimal blocklength for vehicular networks, the paper~\cite{9465806} for multicast networks under energy constraints, and~\cite{8437671} that demonstrates the effect of blocklength on the violation probabilities of delay and peak AoI. Notably, the latter is extended for VLSF coding in~\cite{8640078}. All of the above works assume instantaneous feedback. Moreover, they focus on equi-length packets, with the exception of~\cite{8640078} which also considers the \textit{average} blocklength in VLSF coding.

Very few papers consider the feedback sequence in retransmission schemes. We note~\cite{9013381}, which draws the AoI-optimal feedback sequence for stationary policies with instantaneous feedback. In~\cite{10001535}, the presence of feedback incurs a reduction in the number of transmissions, leading to insightful results. Our work differentiates itself by directly considering the feedback delay.

To the best of our knowledge, this is the first work to analyze the AoII with VLSF coding. It is also the first to optimize the AoII with positive feedback delay when monitoring sources with small cardinality. Additionally, this is the first time in the AoII research that the feedback sequence is included as part of the optimization problem.

The main contributions of this paper are summarized as follows: 
\begin{itemize}
    \item The average AoII of a Markov source is studied in a VLSF channel coding scheme, where feedback is considered to have a constant (possibly positive) delay.
    \item The optimization problem for the feedback sequence is formulated as a Markov decision process (MDP). We develop MDPs for both AoII-optimal and delay-optimal feedback sequences of VLSF codes. As a baseline reference, we also compute the delay-minimal periodic feedback sequences.
    \item The selected feedback sequences are compared for varying SNR figures. It is demonstrated that a lower average delay does not necessarily correspond to a lower average AoII. On the other hand, the structure of the feedback sequence plays a significant role, exemplified by the surprisingly good performance of the periodic feedback sequences.
\end{itemize}

In Section \ref{sec:Problem}, the key elements of the system are described. Section \ref{sec:Math} presents the mathematical formulation and algorithmic solution of the problem. Numerical results are produced in Section \ref{sec:Results}. In Section \ref{sec:conclusion}, conclusions are made and directions for future work are discussed. 

\section{Problem Definition}\label{sec:Problem}

\subsection{Communication model}\label{sec:Communication}

We consider a discrete-time communication model over a Gaussian channel, where the SNR is denoted by $\gamma$. The duration of one time slot is constant and equal to the transmission time of a single coding symbol, i.e. one channel use. The transmitter monitors a data source and samples it at every time slot. Let the sample at time $t$ be denoted with $X_t$. Note that the samples $X_t$ and $X_{t'}$ might be identical, where we say they carry the same information. At each time-slot $t$, the transmitter takes an action $y_t\in \{0,1\}$, where $y_t=0$ denotes the \emph{wait} action and  $y_t=1$ denotes the \emph{transmit} action. 

We employ variable-length stop-feedback codes, denoted as ($l$,$M$,$\epsilon$) VLSF codes, where $l$ is the average blocklength, $M$ is the codebook size, and $\epsilon$ is the average probability of error~\cite{5961844}. 
More precisely, $l$ is the average number of coding symbols needed for decoding when the decoder makes a decision that can be incorrect with probability $\epsilon$. We assume an incorrect decision outputs a uniformly distributed value over the residual source values.

We focus on the zero-tolerance policy, i.e. transmissions with fresh samples occur as long as the AoII is positive. This policy is a special case of a threshold-based policy where the threshold is equal to zero, and aligns with the results of related works. For example, the works~\cite{bountrogiannis2023age,9137714,kriouile2022minimizing,bib:aoii_semantics1} prove that the optimal transmission policies with resource constraints are threshold-based, where it can be seen that they become the zero-tolerance policies when the resource constraints are taken away.

\subsection{Feedback Sequence}\label{sec:feedback_sequence}
The transmitter sends a predefined number of codeword symbols before receiving a feedback signal. We refer to this group of symbols as a packet, where the first packet is transmitted before the first feedback signal, and so on. The length of the $r$-th packet is fixed and equal to $\nu_r$ for all source values. The sequence $\{\nu_r\}$ is referred to as the \emph{feedback sequence}, as it specifies the times of feedback transmissions. 

Taking into account the transmission time of the feedback signal, an additional delay is incurred which equals a constant $\beta$ number of time slots. A total number of $\nu_r+\beta$ time slots are required for the transmission $r$-th packet and its feedback reception. If the transmitter decides to transmit, the entire $\nu_r$ symbols must be transmitted without interruptions. Accordingly, $\beta$ time slots are required for the feedback signal after transmitting those $\nu_r$ symbols, where the transmitter must stay idle.

Let $p_c(m)\in[0,1)$ denote the probability that the decoder stops (succeeds) at the $m$-th received symbol if decoding is attempted at every symbol. The probability function for the Gaussian channel is estimated according to~\cite{papouvlsf}. When decodings are attempted at each packet, the probability that the decoder succeeds at the $r$-th packet is
\begin{equation}
    p_s(r) \triangleq \sum_{m=L_{r-1}+1}^{L_{r}}p_c(m),
\end{equation}
where $L_{r}\triangleq \sum_{i=1}^{r}\nu_i$ is the total number of symbols received up to the $r$-th packet. Then, the conditional probability $p(r)\in[0,1)$ that the decoder succeeds at the $r$-th packet, given it had failed up to the previous packet is

\begin{equation}\label{eq:p}
   p(r) = \frac{p_s(r)}{1-\sum_{i=1}^{r-1} p_s(i)}. 
\end{equation}
In practical communication systems, the maximum number of packets or channel uses is constrained~\cite{bib:harq_survey}. We achieve this by requiring the following equality,
\begin{equation}
    \sum_i \nu_i = L,
\end{equation}
where $L$ is the maximum number of coding symbols per sample. As $L$ increases, the probability of successful decoding after receiving all packets approaches $1$. For our numerical analysis, we experimentally define $L$ as the maximum number of symbols required for successful decoding in $10^6$ iterations (sample transmissions) of the simulation described in~\cite{papouvlsf}.


\subsection{Source Model}

\begin{figure}
\centering
\vspace{-10pt}
\resizebox{.85\linewidth}{!}{
\begin{tikzpicture}[start chain=going right,scale=0.5]
  \tikzset{%
    in place/.style={
      auto=false,
      fill=white,
      inner sep=2pt,
    },
  }
  \node[state,on chain]                   (0) {$1$};
  \node[state,on chain]                   (1) {$2$};
  \node[state,on chain]                   (2)  {$3$};
  \node[draw=white,fill=white,state,on chain] (3) {$\dots$};
  \node[state,on chain]                   (4)  {$M$};
  \draw[->, auto]
    (0) edge[loop above] node {$\alpha$} ()
    (1) edge[loop above] node {$\alpha$} ()
    (2) edge[loop above] node {$\alpha$} ()
    (4) edge[loop above] node {$\alpha$} ()

    (1) edge[bend right=75] node[pos=0.5, in place] {$\mu$} (2)
    (0) edge[bend right=75] node[pos=0.5, in place] {$\mu$} (1)  	    
    (2) edge[bend right=75] node[pos=0.5, in place] {$\mu$} (4)
   	
	(0) edge[bend right=75] node[pos=0.5, in place] {$\mu$} (2)
	(1) edge[bend right=75] node[pos=0.5, in place] {$\mu$} (4)
	
	(0) edge[bend right=75] node[pos=0.5, in place] {$\mu$} (4)

    (4) edge[bend right=75] node[pos=0.5, in place] {$\mu$} (0)
    (4) edge[bend right=75] node[pos=0.5, in place] {$\mu$} (1)
    (4) edge[bend right=75] node[pos=0.5, in place] {$\mu$} (2)

    (2) edge[bend right=75] node[pos=0.5, in place] {$\mu$} (1)
    (2) edge[bend right=75] node[pos=0.5, in place] {$\mu$} (0)

    (1) edge[bend right=75] node[pos=0.5, in place] {$\mu$} (0)

    ;
\end{tikzpicture}
}
\vspace{-20pt}
\caption{The symmetric Markov data source under consideration.}
\label{fig:source}
\end{figure}
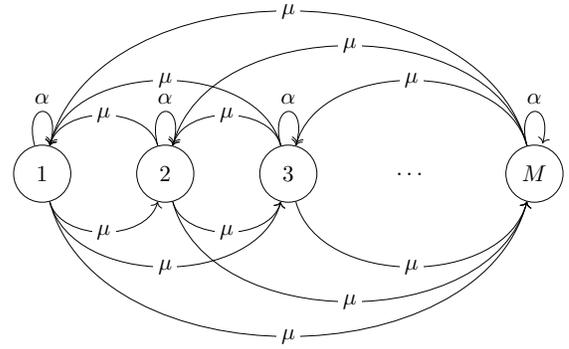

This work focuses on symmetric Markov sources, as illustrated in Figure \ref{fig:source}. In this context, $P(X_{t+1}\!=\!X_t\mid X_t)\!=\!\alpha$, and $P(X_{t+1}\!=\!x\mid X_t)\!=\!\mu\ \forall\ x\in\{1,\dots,M\}\setminus X_t$. This model represents the situation where the source changes every $Y$ time slots, where $Y$ is geometrically distributed with parameter $1-\alpha$, and the source changes uniformly over the residual source values.


Since the cardinality of the source is $M$, each sample requires $k$ bits for its representation, where

\begin{equation}
    k=\lceil \log_2 M \rceil .
\end{equation}

Let $\mathbf{P}$ denote the (single-step) transition probability matrix of the Markov source in Fig.~\ref{fig:source},

\begin{equation}\label{eq:P}
    \mathbf{P}\triangleq
    \bordermatrix{%
    &\scriptstyle\uline{1}&\scriptstyle\uline{2}&\dots&\scriptstyle\uline{M}\cr
    \hfill\scriptstyle\uline{1}\hfill&\alpha &\mu &\dots &\mu\cr
    \hfill\scriptstyle\uline{2}\hfill&\mu &\alpha &\dots &\mu\cr
    \hfill\scriptstyle\vdots\hfill&\vdots&  &\ddots &\vdots\cr
    \hfill\scriptstyle\uline{M}\hfill&\mu &\mu &\dots &\alpha\cr
    }
\end{equation}
The probability of the $t$-th step transition from the state $i$ to the state $j$ is given by the $(i,j)$-th element of $\mathbf{P}^t$, denoted by $\mathbf{P}^t_{ij}$. Due to the symmetry of the Markov chain, $\mathbf{P}^t_{ij}$ is constant and equal for all $i=j$ and $\mathbf{P}^t_{ij}$ is constant and equal for all $i\neq j$.




We assume the following inequality

\begin{equation}\label{eq:P_ineq}
   \mathbf{P}_{ii}^{t} > \mathbf{P}_{ij}^{t},\quad \forall\ i\neq j, t \in \mathbb{N},
\end{equation}
which is equivalent to
\begin{equation}\label{eq:alpha_mu_ineq}
   \alpha > \mu.
\end{equation}
The inequality in~\eqref{eq:P_ineq} implies two important remarks.

\begin{remark}\label{rem:1}The most likely source value during decoding is the transmitted one. Hence, the transmitter sends only the most recent value of the source. Accordingly, if a NACK is received and the source has changed, the transmitter discards the previous sample and sends the most recent.
\end{remark}

\begin{remark}\label{rem:2}The optimal estimator at the receiver outputs the freshest sample available.
\end{remark}

For the error function in~\eqref{eq:aoii_def}, we employ an indicator function,
\begin{equation}\label{eq:error}
    g(X_t,\hat{X_t})=\mathbb{1}_{\{X_t\neq \hat{X}_t\}}\,.
\end{equation}
This error function penalizes any information mismatch between the source and the monitor equally. This function can also arise by truncating more complex error functions. In this case, some information is sacrificed in exchange for the tractability of the optimization problem.

\section{Mathematical Formulation}\label{sec:Math}

This work aims to optimize the feedback sequence $\{\nu_r\}$ for the average AoII. A feedback sequence can be represented as an $L$-length binary sequence, with a $1$ at the $j$-th position indicating feedback after the reception of the $j$-th symbol. Hence, the brute force method for finding the optimal feedback sequence scales in $O(2^L)$. To address this, we utilize an MDP formulation. We assume that an ACK always corresponds to decoding the transmitted value. In practice, this assumption approximates the solution for an $\epsilon$ that is near zero and allows for simpler mathematical derivations. We define an MDP for AoII-optimal and then derive an MDP for delay-optimal feedback sequences.

First, we derive the transition probabilities of the AoII in our system. Due to the indicator error function, we have that

\begin{equation}
    \delta_{t+1}=\begin{cases}
                \delta_t+1 & , \text{ if } X_{t+1}\neq \hat{X}_{t+1}\\
                0 & , \text{ if } X_{t+1} = \hat{X}_{t+1}
    \end{cases}
\end{equation}

Let $D_{t+1}$ denote the event that a successful decoding occurs at time $t\!+\!1$, and $\overline{D}_{t+1}$ the complementary of $D_{t+1}$ (no decoding is attempted or it is deemed unsuccessful). First, we examine the AoII when $\overline{D}_{t+1}$ happens, which means that \mbox{$\hat{X}_{t+1}=\hat{X}_t=\hat{x}$}. 

\begin{figure}
\centering
\begin{tikzpicture}[start chain=going right,transform shape,scale=0.85]
  \tikzset{%
    in place/.style={
      auto=false,
      fill=white,
      inner sep=2pt,
    },
  }
  \node[state,on chain]                   (0) {$0$};
  \node[state,on chain]                   (1) {$1$};
  \node[state,on chain]                   (2)  {$2$};
  \node[state,on chain]                   (3)  {$3$};
  \node[draw=white,fill=white,state,on chain] (4)  {$\dots$};
  \draw[->, auto]
    (0) edge[loop above] node {$\alpha$} ()

    (0) edge[bend left=75] node[pos=0.5, in place] {$1-\alpha$} (1)  	
    (1) edge[bend left=75] node[pos=0.5, in place] {$1-\mu$} (2)    
    (2) edge[bend left=75] node[pos=0.5, in place] {$1-\mu$} (3)
    (3) edge[bend left=75] node[pos=0.5, in place] {$1-\mu$} (4)

    (4) edge[bend left=75] node[pos=0.5, in place] {$\mu$} (0)
    (3) edge[bend left=75] node[pos=0.5, in place] {$\mu$} (0)
    (2) edge[bend left=75] node[pos=0.5, in place] {$\mu$} (0)
    (1) edge[bend left=75] node[pos=0.5, in place] {$\mu$} (0)
    ;
\end{tikzpicture}
\vspace{-20pt}
\caption{The AoII process in the absence of successful decodings.}
\label{fig:AoII_no_decode}
\end{figure}
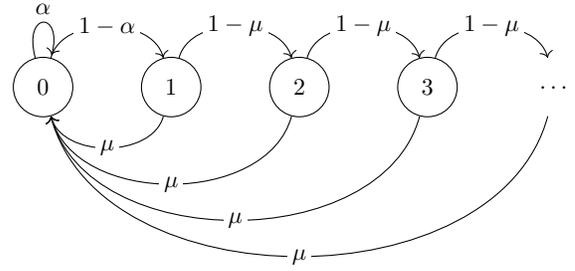

Suppose that $\delta_t=0$, i.e. $X_t=\hat{X}_t$. Then,
\begin{align}
    &P(\delta_{t+1} = 0 \mid \delta_t=0, \overline{D}_{t+1}) = P(X_{t+1} = X_t) = \alpha,\label{eq:d=0_1}\\
    &P(\delta_{t+1} = 1 \mid \delta_t=0, \overline{D}_{t+1}) = P(X_{t+1} \neq X_t) = 1-\alpha.\label{eq:d=0_2}
\end{align}
Next, suppose that $\delta_t>0$, i.e. $X_t\neq\hat{X}_t$,
\begin{align}
    &P(\delta_{t+1} = 0 \mid \delta_t>0, \overline{D}_{t+1}) = P(X_{t+1}=\hat{x}\mid X_t\neq \hat{x}) = \mu,\label{eq:d>0_1}\\
    &P(\delta_{t+1} = \delta_t\!+\!1 \mid \delta_t\!>\!0, \overline{D}_{t+1}) = P(X_{t+1}\!\neq\!\hat{x}\mid X_t\!\neq\! \hat{x}) = 1-\mu.\label{eq:d>0_2}
\end{align}
Thus, in the absence of successful decodings, the AoII progresses as a Markov chain, illustrated in Fig.~\ref{fig:AoII_no_decode}. 

Next, let the event $D_{t+1}$ occur. Let $b_{t+1}$ denote the number of symbols sent for the sample being transmitted at time \mbox{$t+1$}, and also let $l_{t+1}$ denote the corresponding packet length. Note that the transmitter sends only the freshest source values, discarding the old sample and starting a new transmission if needed, and thus the decoded value equals $\hat{X}_{t+1}=X_{t+1-l_{t+1}}$. In this context, the decoded value is \textit{correct} if the source value has not changed, i.e. $X_{t+1}=X_{t+1-l_{t+1}}$, which happens with probability $\mathbf{P}_{ii}^{l_{t+1}}$. Hence,

\begin{align}
    &P(\delta_{t+1} = 0 \mid D_{t+1}) = P(X_{t+1} = X_t) = \mathbf{P}_{ii}^{l_{t+1}},\\
    &P(\delta_{t+1} = \delta_t+1 \mid D_{t+1}) = P(X_{t+1} \neq X_t) = 1-\mathbf{P}_{ii}^{l_{t+1}}\,.
\end{align}

The above imply that only decodings of correct values decrease the AoII. Given this observation, we infer that the optimization problem pertains to the minimization of the time elapsed from the initiation of a transmission until the next successful decoding of a {correct} value.

Notice that the transmitter cannot initiate a new transmission during feedback transmission. Therefore, in the context of the optimization problem, an additional penalty is paid equal to the average AoII between the time of the previous correct value decoding and the subsequent feedback reception. This penalty can be computed utilizing the Markov chain in Fig.~\ref{fig:AoII_no_decode}. Since the feedback delay is equal to $\beta$, it suffices to constrict the length of the Markov chain up to state numbered as $\beta$. Let $\mathbf{B}$ denote the $(\beta\!+\!1)\!\times\!(\beta\!+\!1)$ transition matrix of this Markov chain. Then, the additional penalty due to the feedback delay equals

\begin{equation}
    \begin{aligned}
        d_{\text{feedback}} =& \sum_{t=1}^\beta\sum_{\delta=1}^t \delta\cdot \mathbf{B}_{0\delta}^t\,.
    \end{aligned}
\end{equation}

Given the above, the MDP for the AoII-optimal feedback sequence is defined as follows.
\begin{definition}[AoII-optimal Feedback Sequence MDP]\label{def:AoII_MDP}
The process is an infinite-horizon average-cost MDP, defined as follows.
\begin{itemize}[leftmargin=*]
    \item The state of the MDP at time $t$ is the variable set $S_t=(d_t,b_t,l_t)\in\mathcal{S}$, where $d_t$ is the time spent transmitting without successful decoding, $b_t$ is the total number of symbols sent for the current sample and $l_t$ is the number of symbols sent since the last feedback.
    \item The actions $f_t\in \mathcal{F}$, where the action space $\mathcal{F}=\{0,1\}$ specifies whether feedback is generated ($f_t=1$) or not ($f_t=0$).
    \item The cost of state $S_t$ is $c(S_t)=d_t+d_{\text{feedback}}$. 
    \item To define the transition probability function \mbox{$P(S_{t+1}\mid S_t, f_t)$}, we observe that the \mbox{$j{+}1$-th} packet is transmitted at time $t$ only if a NACK is received and the source has not changed since the previous packet transmission (Remark~\ref{rem:1}), which happens with probability $(1-p(b_t))\mathbf{P}_{ii}^{l_t+\beta}$. When either the source has changed and a NACK is received, happening with probability $p(b_t)(1-\mathbf{P}_{ii}^{l_t+\beta})$, or a successful decoding of an incorrect value occurs, happening with probability $p(b_t)(1-\mathbf{P}_{ii}^{l_t})$, a new sample is transmitted.
    \begin{align*}
            &P((d_t\!+\!1,b_t\!+\!1,l_t\!+\!1) | (d_t,b_t,l_t),0) = 1,\\
            &P((d_t\!+\!\beta,b_t,0) | (d_t,b_t,l_t),1) = (1-p(b_t))\mathbf{P}_{ii}^{l_t+\beta} ,\\
            &P((d_t\!+\!\beta,0,0) | (d_t,b_t,l_t),1) = p(b_t)(1\!-\!\mathbf{P}_{ii}^{l_t})\!+\!(1\!-\!p(b_t))(1\!-\!\mathbf{P}_{ii}^{l_t+\beta}),\\
            &P((0,0,0) | (d_t,b_t,l_t),1) = p(b_t)\mathbf{P}_{ii}^{l_t}.
    \end{align*}
    
\end{itemize}
\end{definition}

Concerning the delay-optimal MDP, we track the transmission time of a single sample. That is, we penalize the time spent until decoding when packets are transmitted without discarding the old sample for a fresher one. For this reason, we do not penalize the delay incurred by the feedback of a previous sample transmission. It can be seen that the delay-optimal MDP is a special case of the AoII-optimal MDP, as follows.
~\\
\begin{definition}[Delay-optimal Feedback Sequence MDP]\label{def:Delay_MDP}
The MDP is similar to the AoII-optimal MDP (Def.~\ref{def:AoII_MDP}), with the exception that $\mathbf{P}_{ii}^{l}\doteq 1\ \forall\ l\in \mathbb{N}$ and $d_{\text{feedback}}\doteq 0$.
\end{definition}
~

The above MDPs are leveraged to derive the optimal feedback sequences as follows. Let $\pi:\mathcal{S}\mapsto\mathcal{F}$ be a \emph{feedback policy}, such that $\pi(S_t)=f_t$. The optimization problem can be expressed as
\begin{equation}
    J_{\pi}(S_0) \triangleq \limsup\limits_{T\rightarrow \infty}\E\left[\left.\frac{1}{T}\sum_{t=0}^{T-1}d_t\right|S_0\right].
\end{equation}

Notice that the MDPs in Def.~\ref{def:AoII_MDP} and Def.~\ref{def:Delay_MDP} are unichain, i.e. there exists a single recurrent class. Hence, by~\cite[Thm. 6.5.2]{bib:krishnamurthy}, the optimal policy $\pi^*$ is independent of the initial state $S_0$ and can be found by solving the following Bellman equations,
\begin{equation}\label{eq:bellman_def}
    g + V(S_t) = \min_{f_t}\{d_t+\sum_{S_{t+1}}P(S_{t+1}\mid S_t, f_t)V(S_{t+1})\},
\end{equation}
where $g \triangleq \min\limits_\pi J_{\pi}(S_0)$ and $V(S_t)$ denotes the \emph{value function} of $S_t$.

An exact solution of the Bellman equations is generally hard to find but can be approximated with the relative value iteration (RVI) algorithm~\cite{bib:krishnamurthy}. Particularly, the RVI approximates the value function iteratively and, when it converges, it finds the true value function. In particular, let $V_\tau(S)$ denote the estimation of $V(S)$ at iteration $\tau\in\mathbb{N}$. Next, define the Bellman operator,
    
    \begin{equation}
        TV_\tau(S) \triangleq \min_{y}\{f(\delta)+\lambda y + \sum_{S'}P(S'\mid S, y)V_\tau(S')\}.
    \end{equation}
    
    Let $S_0$ be some arbitrary reference state. The RVI updates its estimate as follows:
    
    \begin{equation}
        V_{\tau+1}(S) = TV_\tau(S) - TV_\tau(S_0),\quad \forall\ S\in\mathcal{S},\ \tau\in\mathbb{N}\,.
    \end{equation}

Since the RVI encompasses multiple iterations over the state space, it is important to discuss the computational effort due to the matrix power operations. Firstly, the operation $\mathbf{B}^t$, $t=1,\dots,\beta$ is commonly easy to compute since $\mathbf{B}$ is $(\beta\!+\!1)\times(\beta\!+\!1)$ and $\beta$ is typically a small figure. On the other hand, the computation of $\mathbf{P}^l$, $l=1,\dots,L\!+\!\beta$ can be very expensive as $\mathbf{P}$ is $2^k\times 2^k$. To tackle this issue, we develop an efficient version of this operation. The main idea is to split the source states into two groups: the first contains only the current source state, and the second contains the rest. Due to the symmetry of the source model, we do not lose any information by considering the transitions between those two groups. To this end, define the simplified transition matrix
\begin{equation}
    \boldsymbol{\mathcal{P}} \triangleq
    \begin{pmatrix}
    \alpha &1-\alpha\cr
    \mu &1-\mu\cr
    \end{pmatrix},
\end{equation}
and notice that $\mathbf{P}_{ii}^{s}=\boldsymbol{\mathcal{P}}_{11}^{s}$ for any power $s$, reducing dramatically the computational complexity of the power operator.

Having derived the optimal feedback policy for each $(d_t,b_t,l_t)$ MDP state, we shall extract the feedback sequence $\{\nu_r\}$. To this end, beginning at the state $(0,0,0)$, we transit to the next state based on the feedback policy assuming that the source does not change and the decoding fails. This process is repeated until a state with parameter $b_t=L$ is reached. This is described with Algorithm 1.

\begin{algorithm}
    \caption{Extract feedback sequence from MDP solution}
    \label{alg:feedback}
    \begin{algorithmic}[1]
    \renewcommand\algorithmicrequire{\textbf{Input:}}
    \renewcommand\algorithmicensure{\textbf{Output:}}
        \REQUIRE Optimal state-dependent actions $f(d,b,l)$
        \ENSURE Feedback sequence $\{\nu_r\}$
        \\ \textit{Initialization of packet count and MDP state:}
        \STATE $r\gets 1$
        \STATE $(d,b,l)\gets (0,0,0)$
        \\ \textit{Main iteration:}
        \FOR{$b=0,\dots,L$}
            \IF{$y(d,b,l)=0$}
                \STATE $l\gets l+1$
                \STATE $d\gets d+1$
            \ELSE
                \STATE $\nu_r\gets l$
                \STATE $l\gets 0$
                \STATE $d\gets d+\beta$
                \STATE $r\gets r+1$
                \ENDIF
        \ENDFOR
    \end{algorithmic}
\end{algorithm}

Lastly, as a baseline reference, we compare the above feedback sequences with a minimum-delay periodic feedback sequence. A periodic feedback sequence has the property that $\nu_r=\nu^\prime\ \forall\ r$. Due to the periodic property, the brute force method scales in $O(L)$, so an exact solution is easy to find.
~\\

\begin{definition}[Minimum-delay Periodic Feedback Sequence]\label{def:Periodic}
Feedback is transmitted periodically, i.e. $\nu_r=\nu^\prime\ \forall\ r$, where
    \begin{equation}
        \nu^\prime = \argmin_{\nu\in\{1,\dots,L\}} \sum_{r=1}^{\lceil\frac{L}{\nu}\rceil} r(\nu+\beta)p_s(r),
    \end{equation}
    with the convention that $p_s(r)=\sum\limits_{m=L_{r-1}+1}^{L}p_c(m)$ for $r>\frac{L}{\nu}$.
\end{definition}

The last sentence in Def.~\ref{def:Periodic} implies that when $L$ is not divisible by $\nu$, the transmitter stops transmitting the final packet at the $L$-th symbol, whilst feedback is generated when the period ends at the $L_{r-1}+\nu$ slot. Thus, the constraint on the maximum number of transmitted symbols per sample applies without limiting the search space to the periods that divide $L$.

\begin{figure}
\centering
\hspace{9.5pt}\includegraphics[width=0.9504\linewidth]{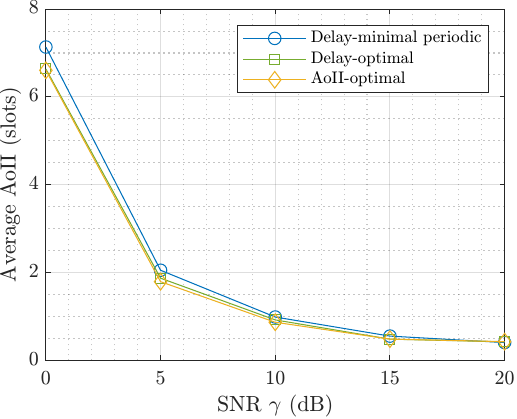}
\vspace{-10pt}
\caption{Average AoII as a function of SNR for feedback delay $\beta=1$ and $k=10$ bits.}
\label{fig:aoii10}
~\\
\hspace{0pt}\includegraphics[width=0.988\linewidth]{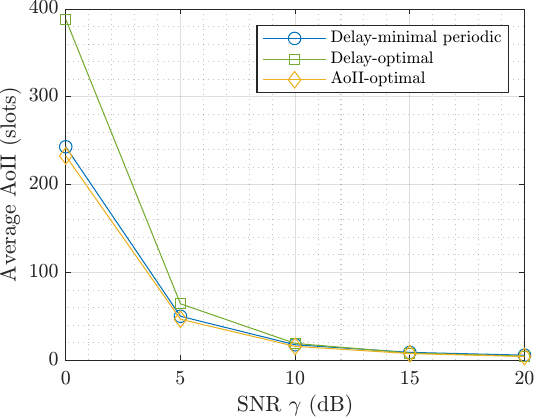}
\vspace{-10pt}
\caption{Average AoII as a function of SNR for feedback delay $\beta=1$ and $k=100$ bits.}
\label{fig:aoii100}
\end{figure}

\begin{figure}
\centering
\hspace{4.5pt}\includegraphics[width=0.9702\linewidth]{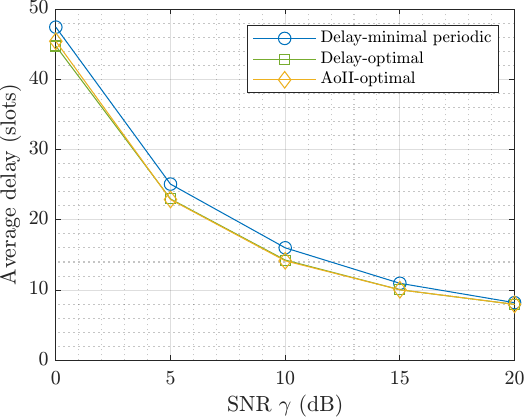}
\vspace{-10pt}
\caption{Average delay as a function of SNR for feedback delay $\beta=1$ and $k=10$ bits.}
\label{fig:delay10}
~\\
\includegraphics[width=0.99\linewidth]{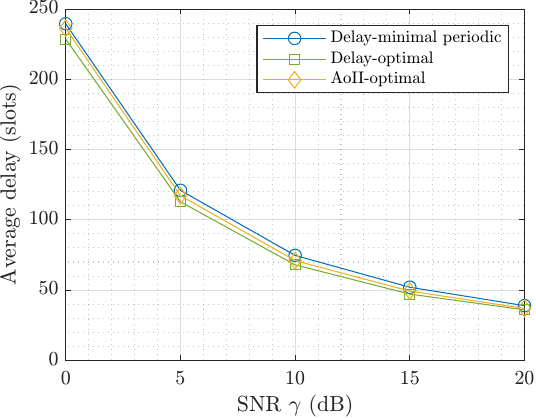}
\vspace{-10pt}
\caption{Average delay as a function of SNR for feedback delay $\beta=1$ and $k\!=\!100$ bits.}
\label{fig:delay100}
\end{figure}

\section{Numerical Results}\label{sec:Results}

\begin{figure}
\centering
\hspace{5pt}\includegraphics[width=.98\linewidth]{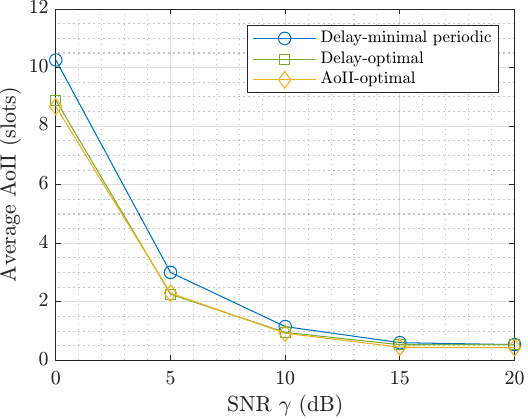}
\vspace{-10pt}
\caption{Average AoII as a function of signal-to-noise ratio for feedback delay $\beta=4$ and $k=10$ bits.}
\label{fig:aoii10_beta4}
~\\~\\
\includegraphics[width=\linewidth]{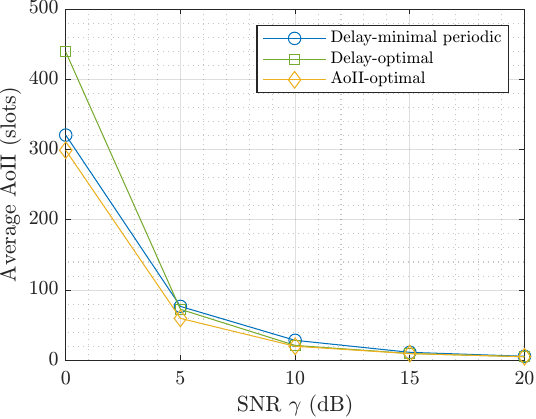}
\vspace{-20pt}
\caption{Average AoII as a function of signal-to-noise ratio for feedback delay $\beta=4$ and $k=100$ bits.}
\label{fig:aoii100_beta4}
\end{figure}

\begin{figure}
\centering
\hspace{5pt}\includegraphics[width=.98\linewidth]{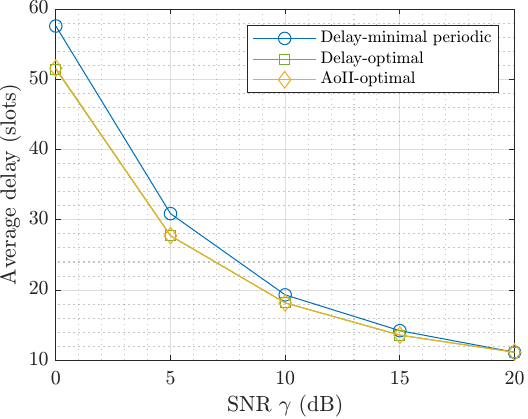}
\vspace{-10pt}
\caption{Average delay as a function of signal-to-noise ratio for feedback delay $\beta=1$ and $k=10$ bits.}
\label{fig:delay10_beta4}
~\\~\\
\includegraphics[width=\linewidth]{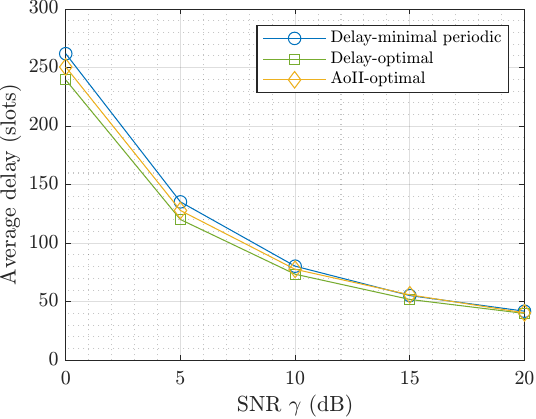}
\vspace{-20pt}
\caption{Average delay as a function of signal-to-noise ratio for feedback delay $\beta=1$ and $k=100$ bits.}
\label{fig:delay100_beta4}
\end{figure}

\begin{figure*}
\centering
    \subfloat[$\gamma=0$]{\includegraphics[width=0.328\linewidth]{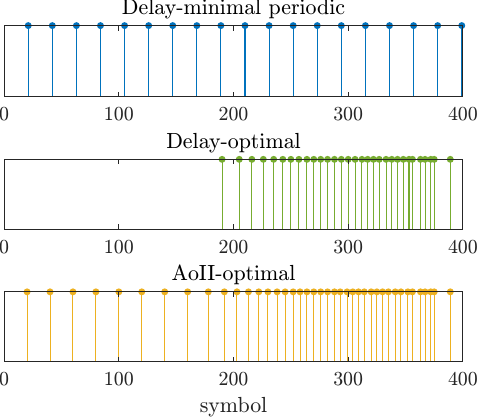}}\,
    \subfloat[$\gamma=5$]{\includegraphics[width=0.328\linewidth]{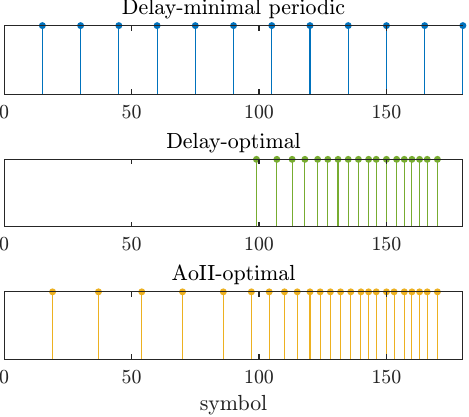}}\,
    \subfloat[$\gamma=15$]{\includegraphics[width=0.328\linewidth]{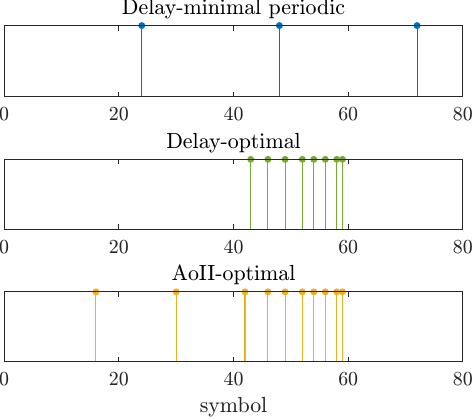}}
\caption{Feedback sequences for $\beta=1$, $k=100$. A spike at the $m$-th symbol indicates that feedback is generated after its reception.}\label{fig:feedback100}
~\\
    \subfloat[$\gamma=0$]{\includegraphics[width=0.328\linewidth]{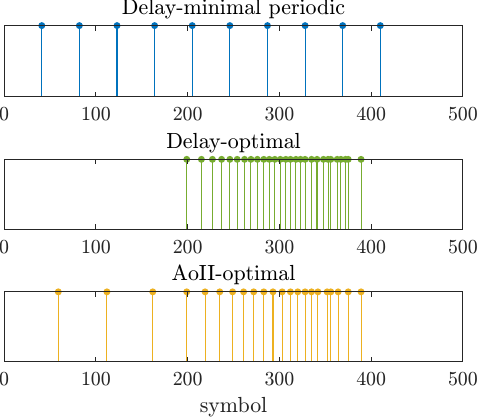}}\,
    \subfloat[$\gamma=5$]{\includegraphics[width=0.328\linewidth]{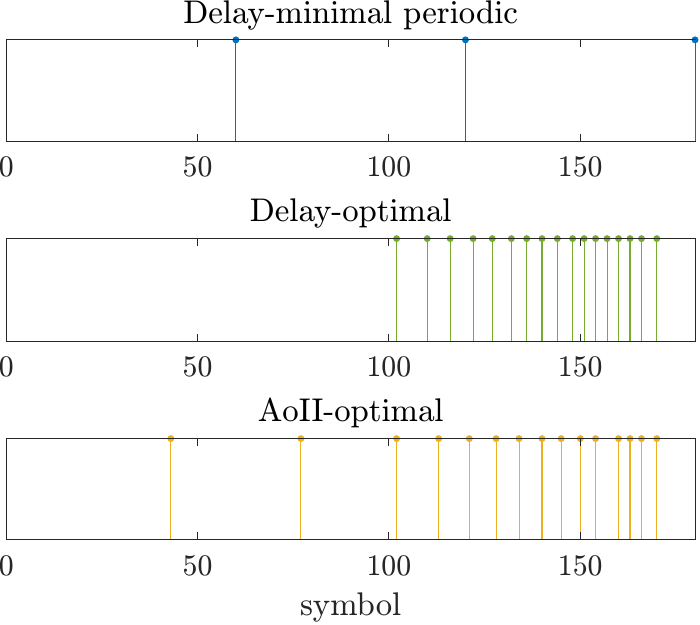}}\,
    \subfloat[$\gamma=15$]{\includegraphics[width=0.328\linewidth]{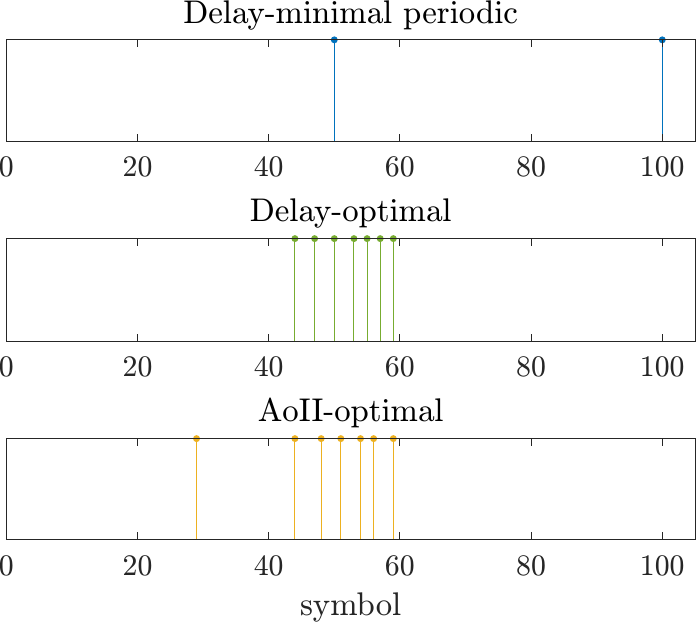}}
\caption{Feedback sequences for $\beta=4$, $k=100$. A spike at the $m$-th symbol indicates that feedback is generated after its reception.}\label{fig:feedback100_beta4}
\end{figure*}

In this section, numerical results are obtained by performing $1000$ Monte Carlo experiments of horizon equal to $T=10^5$. The chosen parameters for the experiments are $\alpha=0.995$, $k\in\{10,100\}$, average probability of error $\epsilon=10^{-3}$, feedback delay $\beta\in \{1,4\}$, and SNR $\gamma\in \{0,5,10,15,20\}$ (dB). Note that $\alpha$ is seemingly large because each time slot equals the transmission time of one coding symbol, which is typically very short. The average AoII for $\beta=1$ is shown in Figures~\ref{fig:aoii10}-\ref{fig:aoii100}, for $k=10$ and $k=100$ respectively, whereas Figures~\ref{fig:delay10}-\ref{fig:delay100} illustrate the delay. Corresponding results for $\beta=4$ are shown in Figures~\ref{fig:aoii10_beta4}-\ref{fig:delay100_beta4}.

We observe that the delay-optimal and AoII optimal feedback sequences do not coincide, while their performance difference increases with $k$. Notably, the delay-optimal feedback does not always perform better than the conventional periodic feedback. This is apparent for $k=100$ and $\gamma<10$. A potential explanation for this can be given by examining the feedback sequences, shown in Fig.~\ref{fig:feedback100} for $\beta=1,\ \gamma\in\{0,5,15\}$ and in Fig.~\ref{fig:feedback100_beta4} for $\beta=4,\ \gamma\in\{0,5,15\}$. We conjecture that since decisions are made only after the reception of feedback, the content of a sufficiently long packet becomes obsolete with a high probability before it is attempted to be decoded, while the transmitter does not have the chance to discard the obsolete sample early. However, this is not the case in high SNRs or small sources because packets are generally short.

\section{Conclusions}\label{sec:conclusion}

This paper focuses on the minimization of the AoII using VLSF coding, where we studied the role of the time instances of feedback generation, i.e. the feedback sequence, when feedback delay is positive. Assuming a Markov source, we formulated the problem as an MDP and derived AoII-optimal and delay-optimal feedback sequences. As a baseline reference, we employed delay-minimal periodic feedback sequences. Numerical results illustrate that delay-optimality does not necessarily imply AoII-optimality. Significantly, periodic feedback sequences perform consistently close to the AoII-optimal, In contrast, delay-optimal sequences may deviate substantially, showing that the structure of the feedback sequence plays a significant role as it allows the scheduler to not only stop the transmission early but also make a new decision. Following this observation, future work could explore efficient methods for deriving AoII-minimal periodic sequences. Notably, periodic feedback is easy to implement in general networks, rendering it attractive for various applications. 

\begin{acks}
This work was partially funded by the project POLAR (FORTH Synergy Grant 2021) and by the European Union -- NextGenerationEU under the framework of the National Recovery and Resilience Plan Greece 2.0, grant (Project: Smart Cities, TAEDR-0536642), and partially by the Engineering and Physical Sciences Research Council (EP/L016656/1) and the University of Bristol.
\end{acks}

\clearpage
\bibliographystyle{ACM-Reference-Format}
\bibliography{biblio}

\end{document}